\documentclass[twocolumn,showpacs,preprintnumbers,amsmath,amssymb]{revtex4}

\usepackage{graphicx}
\usepackage{dcolumn}
\usepackage{bm}

\begin{document}

\preprint{APS/123-QED}


%

\title{Energy partition in low energy fission}
\author{M. Mirea}

\affiliation{Horia Hulubei National Institute for Physics and Nuclear
Engineering, 077125 Bucharest, Romania}

\begin{abstract}
The intrinsic excitation energy of fission fragments is 
dynamically evaluated in terms of the 
time dependent pairing equations. These
equations are corroborated with two conditions. One of them
fixes the number of particles and the another
separates the pairing active spaces associated to the two 
fragments in the vicinity
of the scission configuration. The fission path is obtained in the
frame of the macroscopic-microscopic model.
The single particle level schemes
are obtained within the two center Woods-Saxon shell model.
It is shown that the available intrinsic dissipated energy 
is not shared proportionally to the
masses of the two fission fragments. 
If the heavy fragment possesses nucleon
numbers close to the magic ones, the accumulated intrinsic 
excitation energy is lower than that of the light fragment.
\end{abstract}


\maketitle

\section{Introduction}

Under the action of a mutual Coulomb repulsion, at scission 
the fission fragments are accelerated in opposite directions.
These fragments are 
highly excited, as underlined in many review papers
 \cite{grant,rwh}. The maximal kinetic energy issued in the process
amounts to the $Q$-value in the case of cold fission. 
The fragments decay on their ground state mainly
by evaporation of neutrons and by radiation emission. 
It is known that the motion
of any physical system is governed by conservative forces and
by frictional ones that give rise to dissipation. Therefore, the
excitation energy of the fragments
must depend on the dynamics of the nuclear
system in its path to scission. 

The fission process offers a 
possibility to investigate how two nuclei in contact
share their excitation energy. In the analyzes of experimental
data, the authors of Ref. \cite{sch1} evidenced an energy sorting
mechanism based on statistical arguments. Considering a postulated
difference between the temperatures of the two nascent fragments
in conjunction with the condition of maximum entropy, they emphasized
a flow of energy from one fragment to another. This flow of energy
depends on the available states in the fragments in the scission
configuration. In this context, an explanation for the violation of
the constant temperature hypothesis \cite{bet} that
involves a proportionality between the intrinsic excitation
energy division and the mass ration of the fragment is offered.

Experimental direct indications about the excitation energies of
the fragments are obtained by measuring their 
evaporated neutrons \cite{bow}. Despite a similar temperature
of the neutron velocity distributions, the experiment revealed 
that a larger excitation energy characterizes the light mass
distribution in comparison with the heavy one. The shifting
in the sawtoothlike behavior
of the neutron multiplicity as function of the parent excitation
were attributed mainly to the deformation energy, and not to
the intrinsic heat.
The thermal neutron induced
fission of $^{233}$U \cite{apa} analyzed in Ref. \cite{bur}
evidenced a small neutron multiplicity in the $A$=132 region.
By increasing the excitation energy of the compound nucleus, it
was observed that in this mass region the kinetic energy decreases 
by about 2 MeV, that is a large value.
The interpretation ascribed a similar temperature of fragments
as scission and it was speculated   
that this increment in excitation energy is due to a modification 
of the shape sequences during fission leading to a deformed
heavy fragment. Furthermore, the multiplicity
obtained for two neutron induced $^{237}$Np fission energies \cite{na} 
revealed a modification of the heat of only the fragments in
the heavier mass distribution. It is an experimental indication for
a sorting mechanism in the intrinsic excitation energy.

Motivated by these aspects, in this work,
the intrinsic excitation energy of the fragments 
are evaluated dynamically
in terms of the time dependent 
pairing equations
in the
cold fission regime.
 The macroscopic-microscopic is employed
to obtain the fission path using the minimal action
principle. The method 
is briefly described in the next section.
The basic ingredients for the time dependent pairing equations
 are the single-particle diagrams that must be computed
from the initial state of the fission nucleus up the configuration
given by two separated fragments. The Woods-Saxon
two center shell model \cite{mir08} used to determine realistic
level scheme along the fission path is presented in section \ref{s3}.
In the section \ref{s4} the formalism concerning the time dependent
pairing equations is introduced and its relevance in calculating
dissipation energy is emphasized. 
In section \ref{s5}, the formalism is extended for
two separated nuclei. In Section  \ref{s6} the results concerning the
$^{234}$U fission are reported. The last section is devoted to
conclusions.

\section{The fission trajectory}
\label{s2}

In order to calculate the energy levels diagrams for the fissioning
system, the first step is the determination of a fission path that
satisfies the minimal action criteria \cite{hill}.
The sequence of shapes that follow a nucleus when it passes from
the ground state to the scission point depend principally on
the potential energy surface and the inertia.

In the macroscopic-microscopic method, the whole system is
characterized by some collective coordinates that determine
approximately the behavior of many other intrinsic variables.
The basic ingredient in such an analyzis is the shape
parametrization that depends on several macroscopic degrees of
freedom. The generalized coordinates associated with these degrees
of freedom vary in time leading to a split of the nuclear system
in two separated fragments. The macroscopic deformation energy is
calculated within the liquid drop model. A microscopic potential
must be constructed to be consistent with this nuclear shape
parametrization. A microscopic correction is then evaluated using
the Strutinsky procedure.

First of all, in our description,
it is required to define a nuclear shape 
parametrization. In the
following,
an axial symmetric nuclear shape 
is
obtained by smoothly joining two spheroids of semi-axis $a_i$ and $b_i$
($i$=1,2) with a neck surface generated by the rotation of a circle
around the axis of symmetry. By imposing the condition of volume
conservation we are left by five independent generalized
coordinates \{$q_i$\} ($i$=1,5)
that can be associated to five degrees of freedom: the
elongation $R$ given by the distance between the centers of the
spheroids; the necking parameter $C=S/R_3$ related to the curvature of the
neck, the eccentricities $\epsilon_i$ associated with the deformations of the
nascent fragments and the mass asymmetry parameter $\eta=a_1/a_2$. The notations
that describe this parametrization can be identified by inspecting
the Fig. 1. Due to the axial symmetry, the surface equation is
given in cylindrical coordinates for the three regions involved:
\begin{equation}
\rho(z)=\left\{\begin{array}{lr}b_1\sqrt{1-(z-z_1)^2/a_1^2},& z\le z_{c1};\\
\rho_3-S\sqrt{R_3^2-(z-z_3)^2},&  z_{c1}<z<z_{c2};\\
b_2\sqrt{1-(z-z_2)^2/a_2^2},& z\ge z_{c2}.\end{array}\right.
\end{equation}
It is known that a nuclear shape is well adapted for the fission
process is the following conditions are satisfied \cite{c6}:
(i) The three most important degrees of freedom, that is,
elongation, necking and mass-asymmetry are taken into account;
(ii) A single sphere and two separated fragments are allowed
configurations; (iii) The flatness of the neck is an independent
variable. All these conditions are fulfilled by the above
parametrization.
If $S$=1, the shapes are necked in the median surface characterizing
scission shapes and for $S$=-1 the shapes are swollen addressing the
ground state and the saddle configurations.

\begin{figure}
\includegraphics[width=0.4\textwidth]{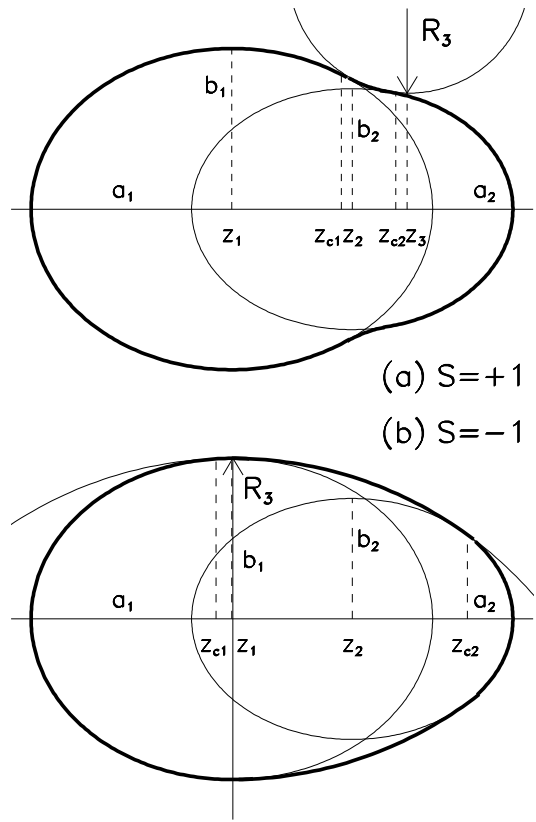}
\caption{Nuclear shape parametrization. Two ellipsoids of different
eccentricities are smoothly joined with a third surface. Two cases
are obtained: (a) the curvature of the neck is positive $S$=1 and
(b) the curvature is negative $S$=-1.\label{fig1}}
\end{figure}

If we consider that the elongation $q_1=R$ is the
main coordinate, the dependencies of the other generalized
coordinates $q_{i}=f_i(R)$ ($i=2,5$) must be obtained.
As specified in Ref. \cite{hill}, such
trajectories emerge by minimizing the action functional. 
\begin{equation}
P=-{2\over\hbar}\int_{R_i}^{R_f}\sqrt{2M(q_i,\partial q_i/\partial R)V(q_i)}dR
\label{wkb}
\end{equation}
where $M(q_i,\partial q_i/\partial R)$ is the inertia along the trajectory and 
$V(q_i)$ is the deformation energy. $R_i$ and $R_f$ stand for the elongation
associated to the ground state and the exit from the barrier, respectively.
In our calculation the reference of the deformation
energy is always taken as the energy in the ground state. So the next condition
is fulfilled $V(R_i)=V(R_f)=0$. As it can be seen in formula (\ref{wkb}),
as the fissioning nucleus passes from its ground state to the
scission configuration, the sequences of shapes depends mainly on
the deformation energy and the inertia.
The deformation energy is obtained in the frame of the macroscopic-microscopic 
model \cite{c4} while the inertia is computed within the
cranking approximation \cite{hill,rrp,rjp}.
The deformation energy was obtained by summing the liquid drop
energy $E_{LDM}$ with the shell and the pairing corrections $\delta E$.
\begin{equation}
V=E_{LDM}+\delta E
\end{equation}
The macroscopic energy $E_{LDM}$ is obtained in the framework of the
Yukawa - plus - exponential model \cite{ktr} extended for binary systems
with different charge densities as detailed in Ref. \cite{mm}:
\begin{equation}
E_{LDM}=E_n+E_C+E_V
\end{equation}
where
\begin{equation}
E_n=-{a_2\over 8\pi^2r_0^2a^4}\int_{v_{1}}\int_{v_{2}}\left({r_{12}\over a}-2\right)
{\exp\left(-{r_{12}\over a}\right)\over {r_{12}\over a}}
d^3r_1d^3r_2
\end{equation}
is the nuclear term,
\begin{equation}
E_C={1\over 2}\int_{v_{1}}\int_{v_{2}}{\rho_e(\vec r_1)\rho_e(\vec r_2)\over r_{12}} d^3r_1d^3r_2
\end{equation}
is the Coulomb energy, and $E_V$ is the volume energy. In the
previous definitions $\rho_e$ are charge densities and $r_{12}=|\vec r_1-\vec r_2|$. The
numerical values of the parameters $a_2$, $r_0$, and $a$ are taken from
Ref. \cite{mmol}.

The shell effects $\delta E$ are obtained as a sum between the shell and
the pairing  microscopic corrections. In this context, the
Strutinsky procedure \cite{hill} was used. These corrections represent
the varying parts of the total binding energy caused by the shell
structure. The single particle level diagrams are computed within
the Woods-Saxon superasymmetric two-center shell model. 

The effective mass is computed within the cranking adiabatic
approximation \cite{hill,rrp,rjp}. In a multidimensional deformation
space, where the nuclear shape is described by the set of $n$
independent generalized coordinates {$q_i$}, the inertia
tensor $M_{ij}$ is defined by the equation of the kinetic energy $T$:
\begin{equation}
T={1\over 2}\sum_{i,j=1}^{n}M_{ij}(q_1,...,q_n)
{\partial q_i\over \partial t}{\partial q_j\over \partial t}
\end{equation}
In the adiabatic description of the collective behavior of a nucleus, the nucleons are 
assumed to move in a average deformed potential. Using a Hamiltonian $H(q_1,...q_n)$ that 
includes pairing interactions, introducing the collective parameters {$q_i$} by means of the 
Lagrange multipliers, it is possible to obtain the response of the nuclear system for slow 
changes of the shape within the cranking model formula
\begin{equation}
\begin{array}{c}
M_{ij}(q_1,...,q_n)={2\over\hbar^2}
\sum_{\nu,\mu}{
\left<\mu\left|{\partial H\over\partial q_i}\right|\nu\right>
\left<\nu\left|{\partial H\over\partial q_j}\right|\mu\right>\over
(E_{\mu}+E_{\nu})^3}\\
\times
(u_\mu v_\nu+u_\nu v_\mu)^2+P_{ij}
\end{array}
\label{cm}
\end{equation}
where $|\nu>$ and $|\mu>$ are single particle wave functions, $E_\nu$, $u_\nu$ 
and $v_\nu$ are the quasiparticle energy, the vacancy and occupation amplitudes of 
the state $\nu$, respectively, in the BCS approximation, and $P_{ij}$ is a correction 
that depends on the variation of the pairing gap and the Fermi energy as function 
of the deformation coordinates. Recently, the formula (\ref{cm}) was generalized by taking 
into account the intrinsic excitation produced during the fission process 
itself \cite{jpg}. The inertia $M$ along a trajectory in the configuration space 
spanned by the generalized coordinates $q_i$ ($i$=1,5) can be obtained within the 
formula
\begin{equation}
M=\sum_{i=1}^5\sum_{j=1}^5M_{ij}{\partial q_{i}\over\partial R}
{\partial q_{j}\over\partial R}
\end{equation}
The  total inertia is the sum of the contributions that correspond
to   the  proton  and  to  the  neutron  level  schemes.  
Usually,  the  matrix
elements of the derivatives of the Hamiltonian in Rel. (\ref{cm})
are replaced by the
matrix  elements  of  the derivatives of the mean field potential alone.

\section{Single particle energies}
\label{s3}

A microscopic potential must be constructed to be consistent
within our nuclear shape parametrization. The simplest way
it to use a semi-phenomenological Woods-Saxon potential. In order
to take into account nuclear deformations going over to
separate shapes and obtain two separated fragments,
a two-center shell model with a Woods-Saxon 
potential was
developed recently \cite{mir08}. Other recipes that allows
to treat strongly deformed nuclei are presented in Ref. \cite{sama,dt}.
The mean field potential is defined in the frame of the Woods-
Saxon model:
\begin{equation}
V_{0}(\rho,z)=-{V_{c}\over 1+\exp\left[{\Delta(\rho,z)\over
a}\right]}
\end{equation}
where $\Delta(\rho,z)$ represents the distance
between a point $(\rho,z)$
and the nuclear surface. This distance is measured only along
the normal direction on the surface and it is negative if the
point $(\rho,z)$ is located
in the
interior of the nucleus. $V_{c}$ is the depth of the
potential while $a$ is the diffuseness parameter. In our work, the
depth is $V_{c}=V_{0c}[1\pm \kappa (N_{0}-Z_{0})/N_{0}+Z_{0})]$
with plus sign for
protons and minus sign for neutrons, $V_{0c}$= 51 MeV, $a$=0.67
fm,
$\kappa$=0.67.  Here $A_{0}$, $N_{0}$ and $Z_{0}$ represent the
mass number, the neutron number and the charge number of the
parent, respectively.
This parametrization, referred as the Blomqvist-Walhlborn one
in Ref. \cite{scwiok}, is adopted
because it provides the same
radius constant $r_{0}$ for the mean field and the pairing field.
That ensures a consistency of the shapes of the two fields at
hyperdeformations, i.e., two tangent ellipsoids.
The Hamiltonian is obtained by adding the spin-orbit and
the Coulomb terms to the Woods-Saxon potential. The eigenvalues are
obtained by diagonalization of the Hamiltonian in the semi-symmetric
harmonic two center basis \cite{mar,mirtc,mirtc2}. In this work, the
major quantum number used is $N_{max}$=12. The two center Woods-Saxon
model will be used to compute shell and pairing corrections together with inertia
in this work. The two center shell model represents a valuable
instrument to investigate the role of individual orbitals for the treatment of a wide variety of nuclear processes 
like cold fission \cite{prc10}, formation of superheavy
elements \cite{epl9} or superasymmetric 
disintegration processes, pertaining to 
cluster- and alpha-decays \cite{m1,m2,m3}.

\section{Time Dependent Pairing Equations}
\label{s4}

In the actual formalism, the starting point is
a many-body 
Hamiltonian with pairing residual interactions. This Hamiltonian
 depends on some time-dependent collective parameters
$q(t)=\{q_{i}(t)\}$ ($i=1,...n$), such as the inter-nuclear 
distances between atoms or nuclei: 
\begin{equation}
H(t)=\sum_{k>0} \epsilon_{k}[q(t)](a_{k}^{+}a_{k}+a_{\bar k}^{+}a_{\bar k})-G\sum_{k,i>0}a_{k}^{+}a_{\bar{k}}^{+}
a_{i}a_{\bar{i}}.
\label{ham1}
\end{equation}
Here, $\epsilon_{k}$ are single-particle energies of the molecular potential,
 $a_{k}^{+}$ and $a_{k}$ denote operators for creating
and destroying a particle in the state $k$, respectively. The state characterized
by a bar signifies the time-reversed partner of a pair.
The pairing correlation arise from the short range interaction 
between fermions moving in time-reversed orbits. The
essential feature of the pairing correction can be described in terms
of a constant pairing interaction $G$ acting between a given number of
 particles. In this paper, the sum over pairs generally runs over
the index $k$. Because the pairing equations diverge for
an infinite number of levels, a limited number of levels are used
in the calculation, that is $N$ levels above and below the
Fermi energy $E_{F}$.

In order to obtain the equations of motion, we shall start from the variational
principle taking the following energy functional
\begin{eqnarray}
{\cal{L}}=<\varphi \mid H-i\hbar{\partial\over \partial t}\mid \varphi >
\label{expr0}
\end{eqnarray}
by assuming the many-body state as time dependent BCS seniority-zero wave function
\begin{eqnarray}
\mid\varphi(t)>&=&\prod_{k}
\left(u_{k}(t)+v_{k}(t)a_{k}^{+}a_{\bar{k}}^{+}\right)
\label{wf2}
\end{eqnarray}
To minimize this functional, the
expression (\ref{expr0}) is derived with respect the independent
variables $v_{k}$,  together with their complex conjugates,
 and the resulting equations
are set to zero.
Eventually, the next coupled-channel equations are obtained :
\begin{equation}
\begin{array}{l}
i\hbar{\dot \rho}_{k}=\kappa_{k}\Delta^{*}-\kappa_{k}^{*}\Delta\\
i\hbar{\dot \kappa}_{k}=(2\rho_{k}-1)\Delta-2\kappa_{k}\epsilon_{k}
\end{array}
\label{bcs1}
\end{equation}
where $\rho_{k}=\mid v_{k}\mid^{2}$ are occupation probabilities,
$\kappa_{k}=u_{k}^{*}v_{k}$ are pairing moment 
components, and $\Delta=G\sum_{k}\kappa_{k}$ is the pairing gap.
$u_{k}$ and $v_{k}$ are the complex BCS occupation and
vacancy amplitudes. 
The variations of single-particle densities $\rho_{k}$
can be evaluated for different values of the generalized velocities
by solving the previous system of coupled equations as done
in Ref \cite{mnap}. Eqs. (\ref{bcs1}) are  also generically 
known as the
time dependent Hartree-Fock-Bogoliubov 
equations \cite{koonin,blocki}. 
As mentioned
in Ref. \cite{blocki}, a connection with the Landau-Zener effect is included in these
equations. Levels undergo Landau-Zener transitions on virtual levels with
coupling strengths given by the magnitude 
of the gap $\Delta$. Recently, these
equations were generalized to take into account the Landau-Zener effect
in seniority one systems
\cite{mpla,mir08} and the pair breaking mechanism \cite{mir4}.

These time-dependent pairing equations can offer a measure
of the dissipated energy.
The difference between the total
energy value $E$ obtained within the TDHFB equations and $E_{0}$ 
given by the static BCS-equations 
 represents an approximate measure for the
dissipation $E_{D}$:
\begin{equation}
E_{D}=E-E_{0}.
\label{bcs2}
\end{equation}
$E$ is expressed simply in terms of
$\rho_{k}$ and $\kappa_{k}$
\begin{equation}
E=2\sum_{k}\epsilon_{k}\rho_{k}-
G\mid \sum_{k} \kappa_{k} \mid^{2}
-G\sum_{k}\rho_{k}^{2}.
\label{eng}
\end{equation}
$E_{0}$ corresponds to $\rho_{k}^{0}$ and $\kappa_{k}^{0}$ associated to the
lower energy state, that is, obtained from BCS equations.

The sum of single particle densities derivatives of 
Eqs. (\ref{bcs1}) is
\begin{equation}
\begin{array}{c}
i\hbar \sum_k {\dot \rho_k}=
\sum_k\left[\kappa_{k}\Delta^{*}-\kappa_{k}^{*}\Delta\right]\\
={1\over G}\left[\mid \Delta\mid^2-\mid\Delta\mid^2\right]=0
\label{p1}
\end{array}
\end{equation}
This result shows that the sum of the single particle
occupation probabilities 
is a constant quantity as the system evolves in 
time according to Eqs. (\ref{bcs1}), that is, the 
average number
of particles in the pairing active space is a constant of
the motion.

\begin{figure*}
\includegraphics[width=0.8\textwidth]{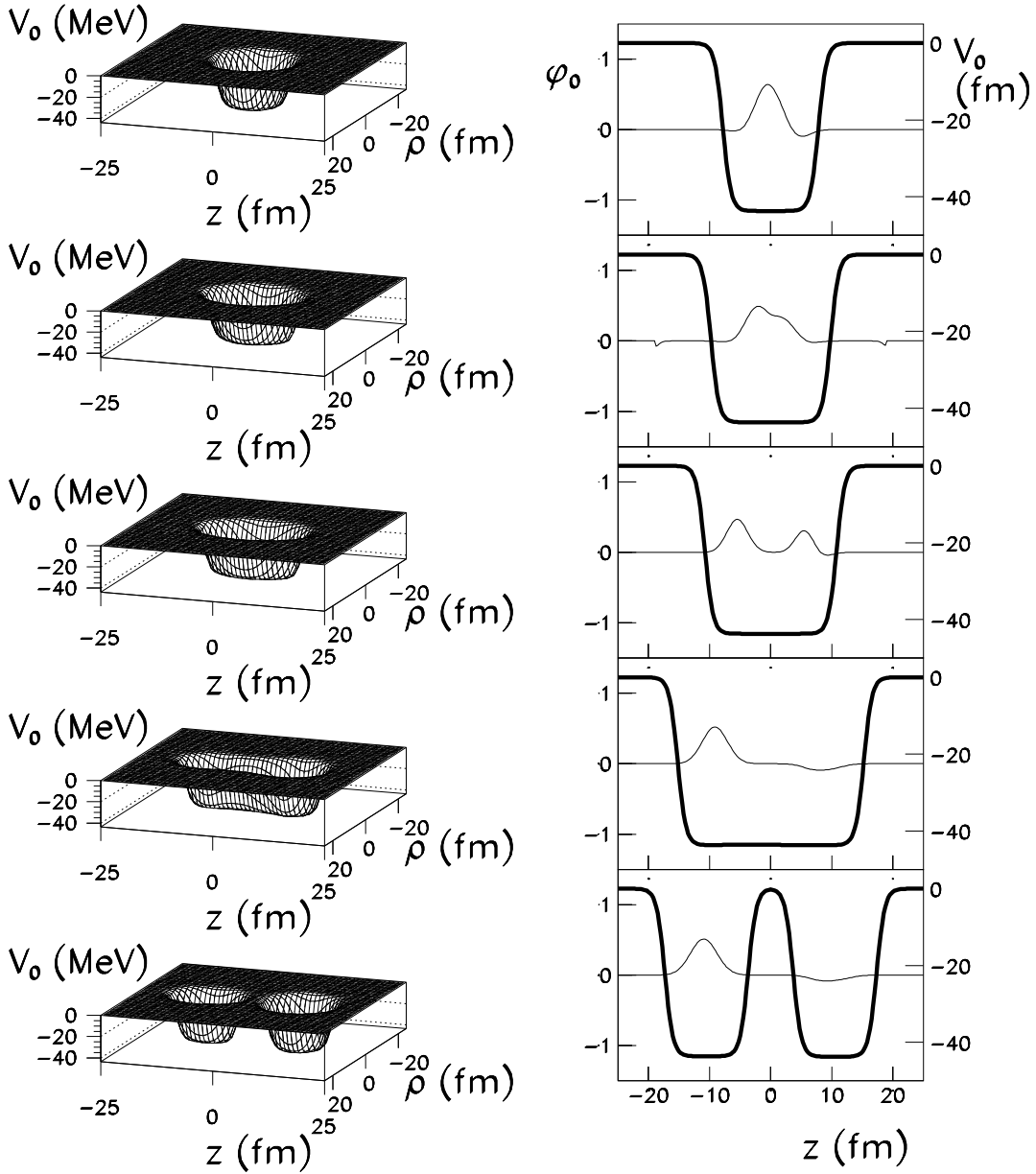}
\caption{Right panel. For the left axis,
the lowest energy Woods-Saxon wave function
for the two center model $\varphi_{0}(z)$ as function of $z$
for different values of the distance between the centers
is displayed with a thin line.
The right axis corresponds to a plot with a thick curve 
of a section in the middle of
the neutron Woods-Saxon potential $V_{0}(z)$  at the same
values of the distance between centers. In the left
panel, a representation of the Woods-Saxon potential $V_{0}$
in the cylindrical $\rho$ and $z$ coordinates is made.
The distances between centers $R$ are 0, 6, 12, 18 and 21 fm from top
to bottom. The configurations displayed correspond to
the minimal  
$^{234}$U fission trajectory.
\label{cig1}}
\end{figure*}

\section{Number of particles}
\label{s5}

After scission, the levels from the pairing active space
will be shared between the two fission fragments. The levels
of the core are sorted accordingly. The sum of occupation 
probabilities of the levels located in each fragment must be
equal to the number of nucleons. This is a problem
that can be solved by appealing to two properties of the
nuclear system.

First of all, the sorting is produced in a continuous manner:
the wave function associated to one single-particle energy
level is transferred gradually to one of the two potential
wells obtained asymptotically, after the scission. For
example, the Woods-Saxon wave function of the lowest single
particle energy is displayed in Fig. \ref{cig1} for different values
of the distance between the centers of the fragments, that is,
for different shapes of the potential. When the two fragments
are completely separated, the wave function is located in one
of the two wells. Making use of this property, it is possible
to identify the final localization of the single particle
level even when the system behave as one single nucleus
if the final mass-asymmetry is known. For this purpose
we calculate two quantities 
$Q_{1k}=<\varphi_k(z)\mid \Theta(-z)\mid \varphi_k(z)>$ and
$Q_{2k}=<\varphi_k(z)\mid \Theta(z)\mid \varphi_k(z)>$ 
where $\Theta$ is the Heaviside function and $\varphi_k$
is the Woods-Saxon single particle wave function of the level $k$ along
the axis $z$. If $Q_{1k}>Q_{2k}$, the wave function is located
in the well 1. It is worth to note that this precedure
fails only in the avoided level crossing regions.

Secondly, the matrix element of the pairing interaction $G$
is in principle dependent of the overlap of the wave functions
of the pairs \cite{delion,ken}.
As long as a single nucleus is involved. the monopole approximation
is considered to perform well \cite{doba} 
and all the values of the matrix elements
are considered to be equal in the active level space.
However, after the scission, the matrix elements of the
pairing interaction between wave function belonging to different
fragments must be zero. If the pairing matrix
elements between pairs located in different fragments at scission
is zero ($G_{12}$=0), then the energy given by Rel. (\ref{eng})
becomes

\begin{equation}
\begin{array}{c}
E=2\sum_{k}\epsilon_{k}\rho_{k}-
G\mid \sum_{k} \kappa_{k} \mid^{2}
-G\sum_{k}\rho_{k}^{2}\\
\rightarrow 2\sum_{k_1}\epsilon_{k_1}\rho_{k_1}-
G_1\mid \sum_{k_1} \kappa_{k_1} \mid^{2}
-G_1\sum_{k_1}\rho_{k_1}^{2}\\
+
\sum_{k_2}\epsilon_{k_2}\rho_{k_2}-
G_2\mid \sum_{k_2} \kappa_{k_2} \mid^{2}
-G_2\sum_{k_2}\rho_{k_2}^{2}\\
=E_1+E_2
\label{eng2}
\end{array}
\end{equation}
where the sum over $k_1$ and $k_2$ means that the
levels belong to fragments 1 and 2, respectively.
We used an arrow in the previous relation to indicate
that the constant value $G$ of the parent nucleus
can be also transformed in the two values $G_1$
and $G_2$ associated to the two fragments.
On other words, the monopole approximation 
is considered valid
separately in each fragment by considering
constant values of $G_1$ and $G_2$.
The relation (\ref{eng2}) shows that the 
total energy $E$
is decomposed in two fragment total energies 
$E_1$ and $E_2$ given by
relations of the type (\ref{eng}) if the
values of the matrix element of the pairing
interaction addressing different fragments
$G_{12}$=0. 

Within these properties, a simple recipe to fix
the average number of particles at scission can be
elaborated. Taking as example the proton level scheme, 
the conditions that the sum of occupation probabilities of
levels in the two wells must be equal with the number of
nucleons of the fragments can be written as:
\begin{equation}
\begin{array}{c}
\label{conditie}
2Z_{p_2}\sum_{k_1}\rho_{k_1}=2Z_{p_1}\sum_{k_2}\rho_{k_2}\\
Z_{p_1}+Z_{c_1}=Z_1\\
Z_{p_2}+Z_{c_2}=Z_2
\end{array}
\end{equation}
where $Z_i$ ($i$=1,2) are the number of protons in the two fragments,
$Z_{c_i}$ and $Z_{p_i}$ stand for the number of protons in the core and 
 the number of protons in the pairing active space, respectively.
For an initial number of pairs $N$ considered in the pairing active
space, the values of $Z_{c_i}$ and $Z_{p_1}$
are simply obtained by counting the levels
given by the two center model. The occupations probabilities
$\rho_{k_i}$ must be obtained from the time dependent pairing
equations (\ref{bcs1}). Exploiting the two previous properties,
the problem to fix the final average 
number of particle within equations
(\ref{bcs1}) is 
now trivial. After the passage of the external saddle point,
in the descent to scission, we insert the 
condition (\ref{conditie}) in the
functional (\ref{expr0}) and we continue to solve the equations
of motion. When the good numbers of particles are obtained,
we impose the condition that $G_{12}$ is zero between the wave functions
belonging to separated fragments. Both conditions (\ref{conditie})
is introduced after the passage of the external saddle point.  
 
In the operator notation the condition (\ref{conditie}) becomes
\begin{equation}
\begin{array}{c}
Z_{p_2}{\hat Z_{p_1}}=Z_{p_1}{\hat Z_{p_2}}\\
{\hat Z_{p_1}}=\sum_{k_1}(a_{k_1}a^{+}_{k_1}+a_{\bar k_1}a^{+}_{\bar k_1})\\
{\hat Z_{p_2}}=\sum_{k_2}(a_{k_2}a^{+}_{k_2}+a_{\bar k_2}a^{+}_{\bar k_2})
\end{array}
\label{co}
\end{equation} 
This condition is introduced in the energy functional (\ref{expr0})
\begin{eqnarray}
{\cal{L}}=<\varphi \mid H-i\hbar{\partial\over \partial t} -\lambda      (Z_{p_2}{\hat Z_{p_1}}-Z_{p_1}{\hat Z_{p_2}})\mid \varphi >
\label{expr1}
\end{eqnarray}
using a Lagrange multiplier $\lambda$. Imposing
also the condition that the interaction matrix element $G$ between
pairs of the same fragment
is not the same than those of different fragments
the new time dependent equations read, eventually:
\begin{equation}
\begin{array}{l}
i\hbar{\dot \rho}_{k_1}=\kappa_{k_1}\Delta_1^{*}-\kappa_{k_1}^{*}\Delta_1\\
i\hbar{\dot \rho}_{k_2}=\kappa_{k_2}\Delta_2^{*}-\kappa_{k_2}^{*}\Delta_2\\
i\hbar{\dot \kappa}_{k_1}=(2\rho_{k_1}-1)\Delta_1-2\kappa_{k_1}(\epsilon_{k_1}-\lambda Z_{p_2})\\
i\hbar{\dot \kappa}_{k_2}=(2\rho_{k_2}-1)\Delta_2-2\kappa_{k_2}(\epsilon_{k_2}+\lambda Z_{p_1})
\end{array}
\label{bcs11}
\end{equation}
where $\Delta_1=G_1\sum_{k_1}\kappa_{k_1}+G_{12}\sum_{k_2}\kappa_{k_2}$
and 
$\Delta_2=G_{12}\sum_{k_1}\kappa_{k_1}+G_{2}\sum_{k_2}\kappa_{k_2}$.
When $G_{12}$=0, it can be easily verified the average 
number of particles in the two fragments are conserved
according to conditions of the type (\ref{p1}) applied separately 
on the two working spaces. That means, $Z_1$ and $Z_2$ behave as
constants. The previous recipe represents the
simplest dynamical
method to project the average number of particles onto two separate
nuclei.

\section{Results}
\label{s6}

To obtain the fission trajectory, 
the action integral (\ref{wkb}) must be minimized in our five-dimensional space.
The first turning point $R_i$ is obtained by determining the ground
state configuration while the second one $R_f$ lies on the
equipotential surface that characterize the exit from the outer
barrier. That means, $R_f$ is defined by the multidimensional function
$V(R,C,\epsilon_1,\epsilon_2,\eta)$=0.  Different methods are currently envisaged to obtain
the heights of the barriers. In static calculations \cite{mol} the
immersion procedure is extensively used while for dynamical paths
\cite{smo,pa,dob} the Ritz method is applicable. To minimize the action
integral we used a numerical method initiated in Ref. \cite{dnp} and used
for fission processes in a large range of mass asymmetries \cite{mir1,mir2,mir3,mir4}
The dependencies $C(R)$, $\epsilon_1(R)$, $\epsilon_2(R)$, and $\eta(R)$ were 
considered as spline functions
of $n$ variables $C_k$, $\epsilon_{1k}$, $\epsilon_{2k}$ and $\eta_{k}$ ($k=1,n$) 
that are associated to the fixed mesh
points $R_k$ located in the interval ($R_i,R_j$). The integral action is
transformed in a numerical function that depends on the 4$n$
variables and it is minimized numerically.

Determination of potential energies $V$ and of effective masses $M_{ij}$ are  very
expensive in computing time. For the numerical minimization procedure,
a large number of iterations is
required and it is not possible to calculate the values of $V$ and $M_{ij}$
for each iteration. An interpolation of calculated
values of the energy and of the effective masses will be used.    
Therefore, to make the problem
tractable, first of all, a grid of deformation values was 
fixed in the five-dimensional configuration space: 25 values of $R$
between 0 fm and 24 fm, 7 values of the eccentricities $\epsilon_i$ between 0
and 0.6, 7 values of the ratio $\eta$ in the interval 1 and 1.6, and 23
values of $C$ between -0.115 fm$^{-1}$ and 0.105 fm$^{-1}$. The pertinent
region of deformations for the possible trajectories between the
ground state and the exit point from the fission barrier was
spanned. In each point of the configuration space, the deformation
energy and the tensor of inertia was computed. During the
minimization process, interpolated values of the deformation energy and
the inertia were used. Different initial values of the generalized 
coordinates were used as input parameters for the minimization.
Therefore different local minimum were obtained. The best value
was selected. Moreover,
additional calculation of the action integral were performed by
slowing varying the generalized coordinates around the best trajectory
previously obtained from the numerical procedure. 
Among all results, the best final trajectory for the least action was 
retained. Such a procedure was used in determining a theoretical
systematic of fission barrier heights in Ref. \cite{cejp}.
In the present work, the trajectory in the configuration space
was modified after the saddle of the outer barrier. The generalized
coordinates were changed to obtain at scission a required 
final configuration
as explained below.

\begin{figure}
\includegraphics[width=0.5\textwidth]{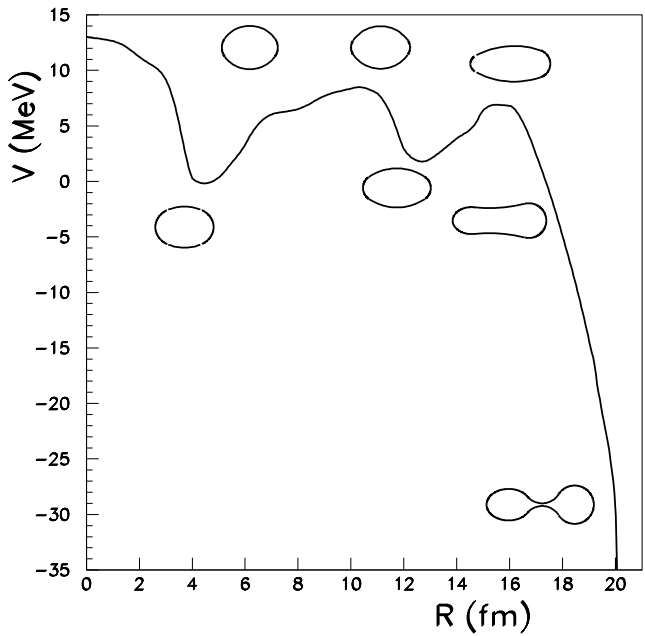}
\caption{
$^{234}$U fission barrier $V$ for a final partition $^{102}$Zr+$^{132}$Te
determined along the minimal action trajectory. Some particular
shapes related to the ground state, the extremes of the barrier, the
exit point and the scission point are inserted in the plot.
The distances for the elongation $R$ that characterizes the shapes
are 4.17, 7.7, 10.43, 12.64, 15.53, 17.53 and 20.2 fm.
\label{bar1}}
\end{figure}

The minimization leads to a definite path in the sub-barrier 
region. Its extrapolation to the scission point is not
unique \cite{ledb}. We will choose the scission configuration
be searching the best candidates that have deformations close
to those obtained in the exit point of the barrier.
We found that the eccentricities and mass-asymmetry parameters 
in the exit point of the outer barrier are 
consistent with the formation
of a pair given by the isotopes $^{102}$Zr and $^{132}$Te.
The ground state configurations of the 
fragments were calculated for this purpose.
By introducing the ground states of the fragments in the
scission configuration, the 
excitation due to the deformations 
energy is minimized and we are left only with the dissipation.
As evidenced from evaluations \cite{wahl}, 
the partition $^{102}$Zr+$^{132}$Te is one of the
larger enought probable partitions. Moreover, this final configuration is also 
interesting due to the fact that the numbers of nucleons 
in the heavy fragments are close to
the magic ones. This configuration allows also a simple comparison
with  the hypothesis made in Ref. \cite{sch1}. The fission barrier
together with some particular shapes are
plotted in Fig. \ref{bar1}.

\begin{figure}
\includegraphics[width=0.5\textwidth]{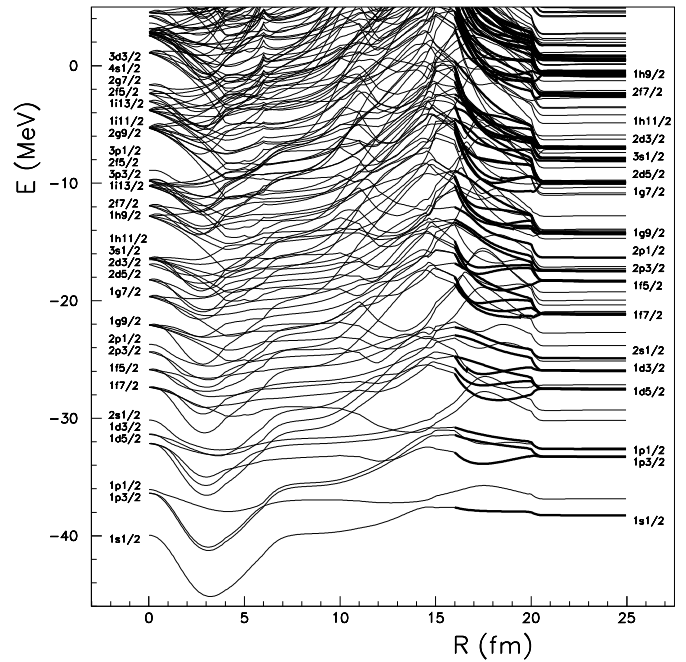}
\caption{ Neutron level diagram for the
$^{234}$U fission with respect the elongation $R$.
\label{ttn}}
\end{figure}

\begin{figure}
\includegraphics[width=0.5\textwidth]{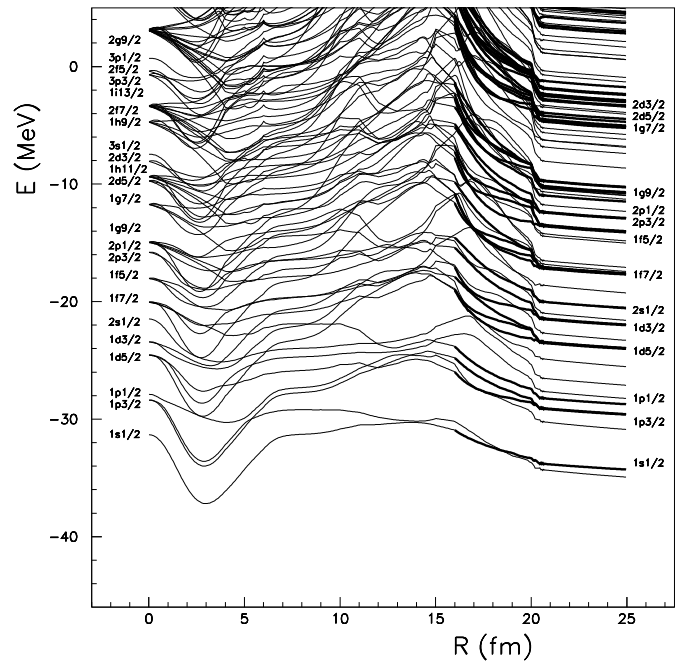}
\caption{ Same as Fig. \ref{ttn} for the
proton diagram.
\label{ttp}}
\end{figure}

The neutron and proton single particle diagrams are calculated
along the minimal action trajectory, from the ground state
of the parent up to the formation of two separated fragments.
These level schemes are plotted in Figs. \ref{ttn} and \ref{ttp}.
In Fig. \ref{ttn}, at $R$=0, the parent nucleus is spherical. For
small deformations the system behavior is
similar to a Nilsson diagram for prolate deformations.
At the right of the figure the orbitals of the 
parent are labeled
with their spectroscopic notations.  
The levels belonging to the heavy fragment
can be identified long before the scission 
configuration and are plotted with thick lines.
The heavy fragment is spherical
at scission, so its levels are bounced in shells.
Making use of the fact that the heavy fragment
becomes spherical after scission, we labeled its shells
with spectroscopic notations. The levels of the light 
fragment are not degenerate due to its deformation.
In the proton diagram of Fig. \ref{ttp}, it can be
observed a smooth decrease of the single particle energies
after the scission due to the Coulomb mutual polarization.
The energy slope for the light fragment is larger than
that for the heavy one.

\begin{figure}
\includegraphics[width=0.5\textwidth]{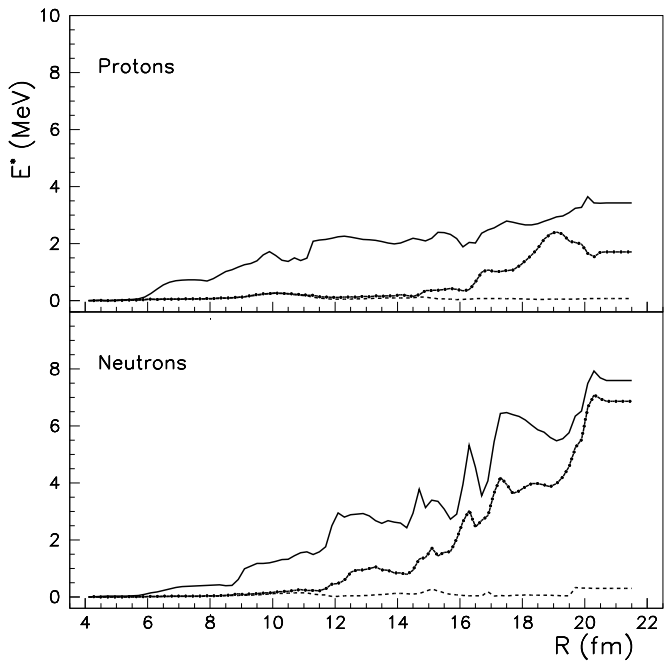}
\caption{ Dissipation energy $E^*$ for protons (upper panel)
and neutrons (lower panel) as function of the elongation $R$.
The calculations are made within Rel. (\ref{bcs2}) and Eqs. (\ref{bcs1})
without imposing any condition for fixing the number of
of particles. The dashed line, the dot dashed line
and the full line  correspond to a inter-nuclear
velocity $dR/dt$ of $10^{4}$, $10^{5}$ and $10^{6}$  m/s,
respectively. 
\label{dis1}}
\end{figure}

It is known that if the projection on a given particle number
changes the pairing coupling constant $G$ \cite{pilat}.
However, for our exploratory investigation, the same $G$
will be kept for the parent nucleus and for the fragments.
After scission, we calculate
the matrix elements for each fragment separately.
The values of $G$, $G_1$, $G_2$ are obtained from the renormalization
procedure \cite{hill} by taking into account the number
of levels in the pairing active space.

To solve the Eqs. (\ref{bcs1}) we need the velocity 
of the generalized
coordinate. 
Our calculations will pertain
to the cold fission regime and are characterized by small values
of the collective kinetic energy and of the excitations energies.
These properties can be obtained by selecting an appropriate
value for the collective velocity.
Therefore, different constant values of the inter-nuclear 
velocity $\dot{R}$ ranging from 10$^{4}$ to
10$^{6}$ m/s were tested. These values can be translated in a time to penetrate
the barrier ranging in the interval $[10^{-18},10^{-20}]$ s.
The time
for the descent between saddle to scission reaches about
4$\times 10^{-21}$ s for a velocity of 10$^{6}$ m/s, a value
considered a typical time for scission \cite{rwh}. 
Within the semi-adiabatic cranking model \cite{jpg}, 
the inertia in the ground state and
in the asymptotic region of two separated fragments are
1.26 and 1.208 $\hbar^2$/MeV$^2$/fm$^2$, respectively. Within
this estimation of the mass, the velocities 
can also be translated in a macroscopic kinetic 
energy that amounts up to 0.3 MeV, of the order of magnitude of
the zero point vibration energy.
The system (\ref{bcs1}) is solved numerically for the
selected velocities. The initial conditions for $\rho_k$ and
$\kappa_k$ were given by the BCS solutions for the parent
ground state, a configuration characterized by $R\approx$4 fm in Fig. \ref{bar1}.
In Fig. \ref{dis1}, the dependencies of the dissipated energy $E^*$ for
the different internuclear velocities investigated 
 are displayed as function of the distance between
the centers of fragments $R$. No conditions for fixing the number
of particles are used. The dissipated energy for the proton
level scheme is lower that that of the neutron one. The dissipated
energy for the lower collective velocity is negligible, while 
for the higher one the total dissipation reaches about 10.5 MeV.
This value is larger than in previous estimations \cite{jpg} because of
the imposed modification of the fission trajectory that gives a particular 
configuration at scission. 
 In connection
with the shape of the barrier displayed in Fig. \ref{bar1}, it can be deduced
that the larger part of the excitation is formed during the penetration
of the second barrier and, in the descend between saddle to fission.
This result is not in line
with the observation that the excitation energy does not increase
after the passage of the saddle point \cite{c5}.

\begin{figure}
\includegraphics[width=0.5\textwidth]{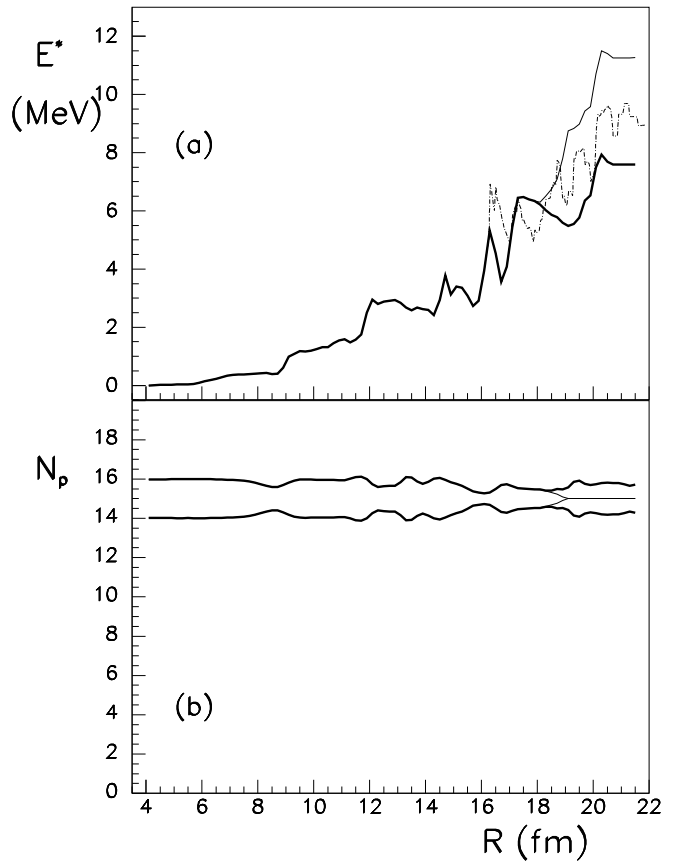}
\caption{
(a) The dissipated energy without constraints
is plotted with a full thick line as function of the
elongation $R$.
The  dissipated energy after imposing the condition for
projecting the number of particles is plotted with a thin line.
The dissipated energy obtained by replacing the ground state
of the parent nucleus within the ground states of the two
nascent fragments.
(b) The number of pairs $N_{p_1}$ and $N_{p_2}$
located on the levels that belongs to
the first and second well are plotted with thick lines.
The number of particles after imposing the conditions for
projecting the number of particles are plotted with thin lines. 
$N_{p_2}$ corresponds to the heavy fragment and it is always larger
than $N_{p_1}$.
\label{excf}}
\end{figure}

It will be interesting to compare the dissipated energy and
the number of nucleons attributed to the two nuclei
in the case of the treatment without constraints and the
treatment with projection of the average number of particle.
In Fig. \ref{excf}(a) the unconstrained dissipation energy
for the neutron workspace is plotted with a thick curve
as function of $R$ at $dR/dt$=10$^6$ m/s. 
The dissipation obtained within the scheme
proposed for projecting the average number of particles
is displayed with a thin line. Up to the the external saddle,
the Eqs. (\ref{bcs1}) are solved . After $R\approx$16 fm, 
we imposed the condition  (\ref{co}).  It can be observed that
the time dependent pairing equations with constraint give
a larger dissipation.  In Fig. \ref{excf}(b) the sums
$N_{p_1}=\sum_{k_1}\rho_{k_1}^2$ and $N_{p_2}=\sum_{k_2}\rho_{k_2}^2$
for the unconstrained equations are plotted with a thick line.
$k_1$ and $k_2$ are levels in the active pairing space
attributed to the fragment 1 and the
fragment 2, respectively. The total number of pairs is conserved
$N_1+N_2=N=30$. After imposing the condition (\ref{co}), the
number of pairs located on the two level schemes of the
two fragment reach the correct values, that is $N_1$=15
and $N_2$=15, that define the partition analyzed.

It was remarked in Ref. \cite{koonin} that the maxima
of the dissipated energy arise because of the character of
the ground state solution is changing rapidly that the
system cannot adjust itself. So, the system appears
excited not because of any changes in the occupation
amplitudes but because the changes in the ground state
to which the dynamic state is studied. 
In this respect,
in Fig. \ref{excf}, we analyzed also the behavior of the system by
replacing the ground state. Instead of the ground state
of the parent, we used as $E_0$ the sum between the 
lower energy states calculated for the two level
schemes of the two nascent fragments. The result is
plotted with a dashed line. At scission we obtain approximately
8 MeV excitation energy for both fragments.

\begin{figure}
\includegraphics[width=0.5\textwidth]{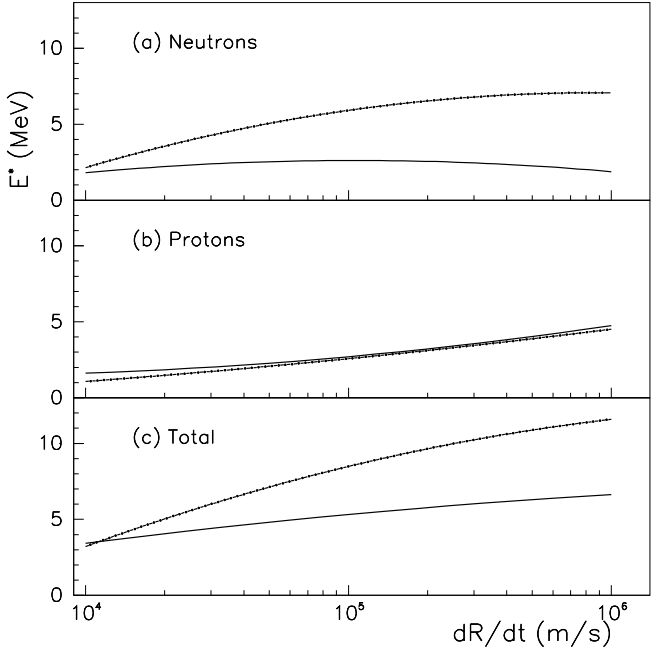}
\caption{
Excitations energies for $^{102}$Zr (dashed line) 
and $^{132}$Te (full line) as
function of the internuclear velocity $dR/dt$. In
panels (a) and (b) the final excitation are displayed
for the neutron and proton schemes, respectively.
In panel (c), the sum between the excitations energies
obtained for protons and neutrons are presented. 
\label{fin1}}
\end{figure}

In Fig. \ref{fin1}, the dissipation obtained for
the two fragments are displayed separately. It is found
that the dissipated energy for the heavy fragment is
much lower than that for the light one. This result is
consistent with the experimental finding addressing the
neutron multiplicities. It is also revealed in Fig. \ref{fin1}
that in the cold regime the increase in excitation
energy of the heavy fragment is smaller than
in the light one, for a modification of the excitation
energy of the parent nucleus. On another hand, the
experimental findings \cite{na} shows that the excitation
energy of the compound nucleus goes almost completely
into excitation energy of the heavy fragments. That means,
a large part of this energy is transferred in deformations
of the heavy nucleus, not taken into account here.
A similar behavior 
is predicted from statistical considerations in Ref. \cite{sch1}
for partitions involving nuclei close to magic numbers.

This model is based on the same philosophical grounds
 as that found 
in Refs. \cite{ledb,ledb2}. The main differences are given
by the time dependent equations used and by the fact that
the dissipation is not evaluated for each fragment
separately. They used the time dependent 
Schr\"odinger equations and introduced quasiparticle excitations
through the cranking approximation.

\section{Conclusions}

Dynamical estimations of the excitation energies in cold fission
are evaluated within time dependent pairing equations. 
Using conditions that fix the number of particles
in each fragment it was possible to obtain for the first time
the excitation
energy of each nucleus issued in the process. A recent
hypothesis \cite{sch1} that claims
the excitation energy is not equilibrated between fragments
was confirmed in the cold fission regime. If the heavy
fragment is close to magic number, its excitation energy
is smaller than that of the light one. It was found that
the dissipation energy in fission fragments is intimately
related on the distribution of pairing occupation probabilities
of the levels at scission. These probabilities can be
obtained by solving an appropriate set of equations of 
motion.

\acknowledgements

 This work was performed in the frame
of the IDEI 512 and MODUL III-EU65 Projects of the Romanian Ministry of Education and Research.

\end{document}